\documentclass[%
 preprint,
superscriptaddress,
 amsmath,amssymb,
 aps,
]{revtex4-1}

\usepackage{graphicx}
\usepackage{dcolumn}
\usepackage{bm}
\usepackage{float}
\usepackage{amssymb}
\usepackage{mathtools}
\usepackage{natbib} 
\usepackage[english]{babel}
\usepackage[utf8]{inputenc}
\usepackage{url}
\usepackage[colorlinks = true,
            linkcolor = blue,
            urlcolor  = blue,
            citecolor = blue,
            anchorcolor = blue]{hyperref}

\begin{document}

\title{Impact of photoexcitation on secondary electron emission: a Monte Carlo study}

\author{Wenkai Ouyang}
 \affiliation{Department of Mechanical Engineering, University of California, Santa Barbara, CA 93106, USA}

\author{Xiangying Zuo}
\affiliation{Department of Mechanical Engineering, University of California, Santa Barbara, CA 93106, USA}

\author{Bolin Liao}
\email{bliao@ucsb.edu} \affiliation{Department of Mechanical Engineering, University of California, Santa Barbara, CA 93106, USA}

\date{\today}

\begin{abstract}
Understanding the transport of photogenerated charge carriers in semiconductors is crucial for applications in photovoltaics, optoelectronics and photo-detectors. While recent experimental studies using scanning ultrafast electron microscopy (SUEM) have demonstrated that the local change in the secondary electron emission induced by photoexcitation enables direct visualization of the photocarrier dynamics in space and time, the origin of the corresponding image contrast still remains unclear. Here, we investigate the impact of photoexcitation on secondary electron emissions from semiconductors using a Monte Carlo simulation aided by time-dependent density functional theory (TDDFT). Particularly, we examine two photo-induced effects: the generation of photocarriers in the sample bulk, and the surface photovoltage (SPV) effect. Using doped silicon as a model system and focusing on primary electron energies below 1 keV, we found that both the hot photocarrier effect immediately after photoexcitation and the SPV effect play dominant roles in changing the secondary electron yield (SEY), while the distribution of photocarriers in the bulk leads to a negligible change in SEY. Our work provides insights into electron-matter interaction under photo-illumiation and paves the way towards a quantitative interpretation of the SUEM contrasts.

\end{abstract}

\maketitle


\section{Introduction}
Understanding photocarrier dynamics accurately is critical for improving the performance and ensuring the stability of high-efficiency photovoltaic (PV) cells, light-emitting diodes, and other optoelectronic devices. For example, the performance of PV cells critically depends on the minority carrier diffusion process \cite{Wilson2020,kim2020high,guo2017long}, and efficient hot photocarrier collection provides a possible means to boost the PV efficiency above the Shockley-Queiser limit\cite{konig2010hot}. These processes can only be probed and better understood by experimental tools that can detect light-induced carrier behaviors with combined high spatial and temporal resolutions. One candidate is the high-energy pulsed electron beam generated by illuminating a photocathode with a pulsed laser source\cite{mourou1982picosecond}. These short electron pulses, with durations down to sub-picoseconds, can be accelerated to high energy, and thus, finely focused to sub-nanometer spatial resolution. Techniques based on the pulsed electron beam, such as ultrafast electron diffraction (UED) and ultrafast transmission electron microscopy (UTEM) have been used to visualize photocarrier dynamics with great success \cite{siwick2003,King2005,zewail2010four,sood2021universal}. In 2010, Zewail and coworkers invented scanning ultrafast electron microscopy (SUEM) by combining the temporal resolution of short electron pulses with the spatial resolution of scanning electron microscopy (SEM)\cite{Yang2010,mohammed20114d,liao2017scanning}. Compared to UED and UTEM, SUEM excels in surface sensitivity and is particularly suitable to study surface photocarrier transport\cite{najafi2015four,liao2017photo, liao2017spatial,ellis2021scanning}, surface acoustic waves\cite{najafi2018imaging,kim2021transient} and surface defect-carrier interactions\cite{shaheen2020real}.

\begin{figure}[htb!]
    \centering
    \includegraphics[width = 0.95\textwidth]{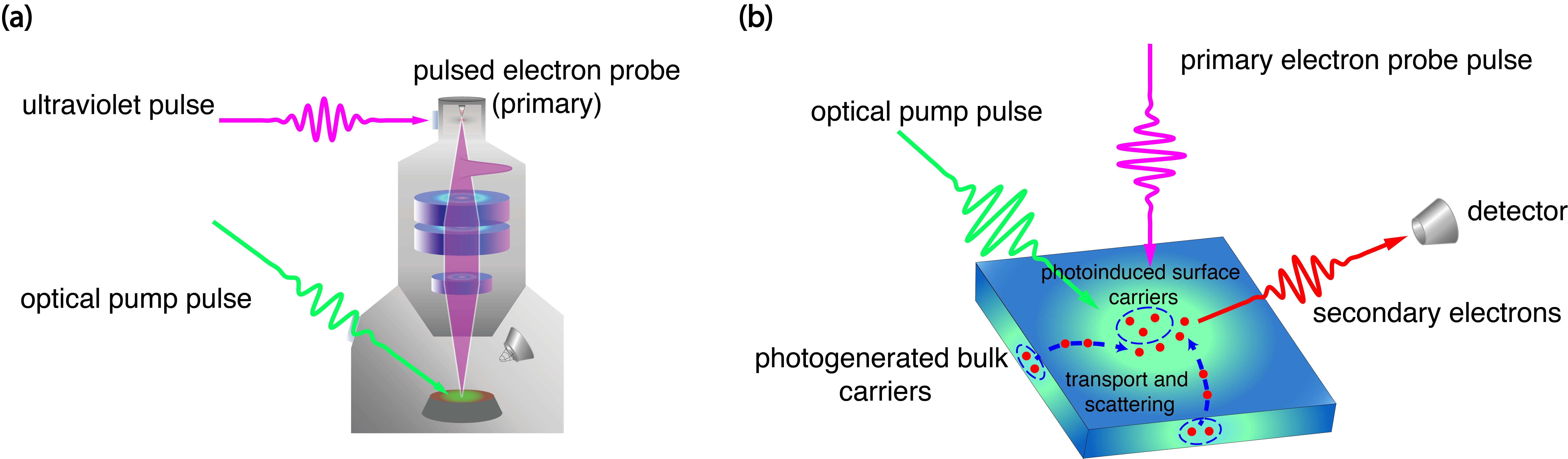}
    \caption{\textbf{Schematic of SUEM experiment and image formation.} (a) A schematic showing a typical setup of the SUEM instrumentation. The green line denotes the optical pump beam while the purple line denotes the optical beam to generate the probe electron pulses. The electron pulses travel down the SEM column and are focused onto the sample through the electron optics. (b) Image formation mechanism in an SUEM. The primary electrons excite secondary electrons from the sample, which are collected by an ETD. The optical pump pulse generates photocarriers in the bulk and near the surface, which modulate the secondary electron yield. The changes in the local secondary electron yield as a result of photoexcitation are used to form SUEM contrast images.}
    \label{fig:1}
\end{figure}

In SUEM, a femtosecond infrared laser source is split and converted into a pump beam (typically 515 nm) and an ultraviolet probe beam (typically 343 nm or 257 nm). Figure \ref{fig:1}(a) depicts the working principle of SUEM. The pump pulse directly hits the sample surface and causes structural, electronic, or thermal  changes, such as photocarrier excitation, surface photovoltages\cite{li2020probing,ellis2021scanning}, topographical distortions and temperature rise\cite{Liao_Najafi_2017}. In the meantime, the ultraviolet probe beam is focused onto the electron gun inside an SEM column to generate short electron pulses, which are accelerated and focused onto the sample surface. The time delay between the optical pump pulses and electronic probe pulses is controlled by a mechanical delay stage. The probe electron pulses, or the ``primary electrons'' (PEs), excite secondary electrons (SEs) from the sample surface,  which are collected by an Everhart-Thornley detector (ETD), and the number of SEs emitted from each location on the sample surface is used to form an image. SUEM contrast images are then generated by comparing the \textit{change} in the SE images as a result of photoexcitation. By controlling the time delay between the optical pump pulses and the electronic probe pulses, SUEM contrast images depicting the change of the sample at a given time after the photoexcitation can be recorded. For example, Najafi et al. observed the interfacial carrier dynamics at a silicon p-n junction with SUEM and reported ballistic carrier transport that cannot be explained by conventional drift-diffusion models\cite{Najafi2017}. Liao et al. measured the hot carrier transport in amorphous silicon using SUEM and observed the spatial separation of electrons and holes\cite{liao2017photo}.   In these studies, the high spatial-temporal resolution of SUEM enables the investigation of photo-induced local microscopic physical processes on a wide range of materials. However, in order to quantitatively interpret these photophysical processes, it is critical to fully understand the underlying mechanisms for the SUEM image contrasts.

\begin{figure}[htb!]
    \centering
    \includegraphics[width = 0.95\textwidth]{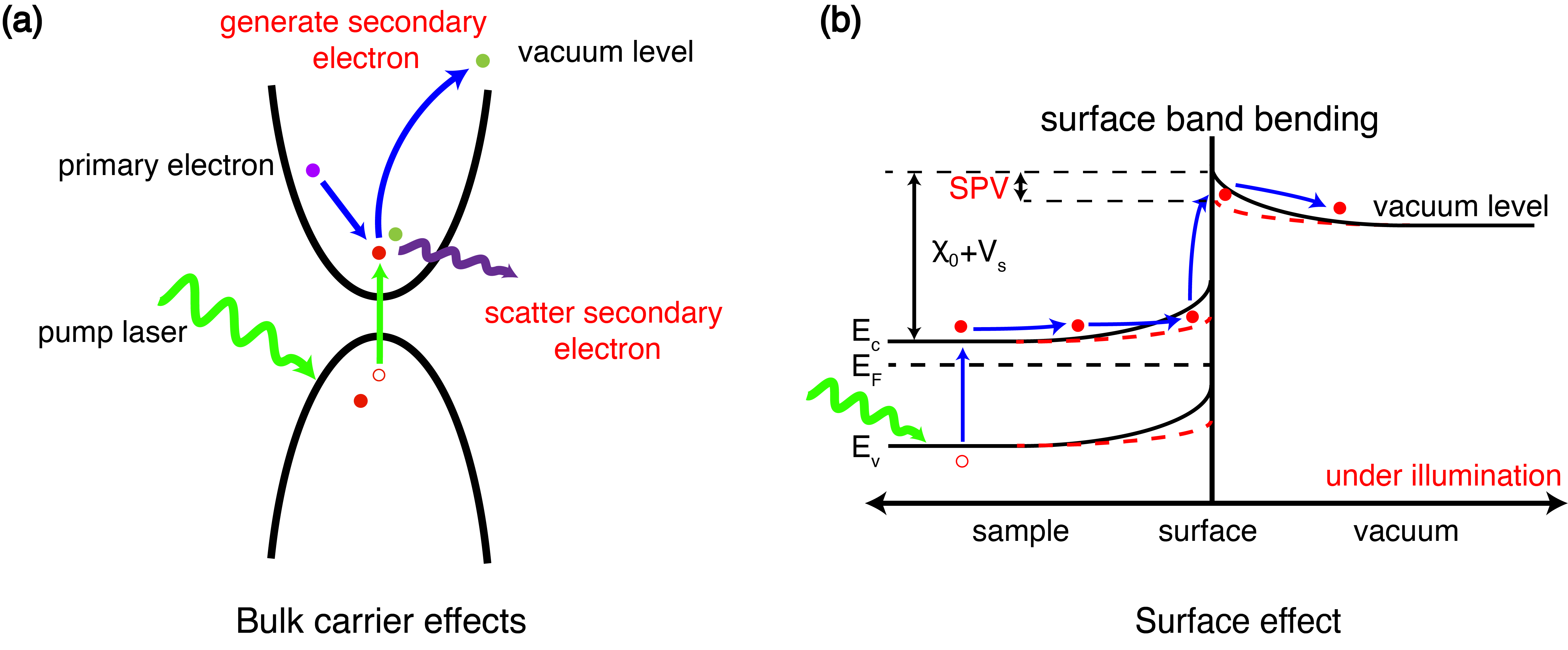}
    \caption{\textbf{Proposed SUEM Contrast Mechanisms.} (a) Schematic illustrating the SUEM contrast mechanism due to the generation of photocarriers in the sample bulk. The presence of photocarriers with higher energy can both increase the generate rate of secondary electrons and scatter the secondary electrons during their transport. (b) Schematic illustrating the surface photovoltage effect in n-type silicon. The black lines depict the energy band diagram in the dark condition, where $\chi_{0}$ is the electron affinity and $V_s$ is the surface band bending due to Fermi level pinning. The red dashed lines depict the energy band diagram under photo-illumination. The photocarriers compensate for the surface band bending and reduces the effective potential barrier for SEs to escape. SPV: surface photovoltage. $E_\mathrm{c}$: conduction band bottom. $E_\mathrm{v}$: valence band top. $E_\mathrm{F}$: Fermi level. }
    \label{fig:2}
\end{figure}

As explained above, SUEM contrast images reflect the change in the number of emitted SEs (SE yield, or SEY) from each location on the sample surface as a result of photoexcitation\cite{liao2017scanning}, as illustrated in Fig.~\ref{fig:1}(b). Thus, the key to understanding SUEM contrast mechanisms is to examine how the SE generation, transport, and emission processes are affected by photoexcitation. Currently, several SUEM contrast mechanisms have been proposed, which can be categorized into bulk effects and surface effects\cite{liao2017scanning}, as summarized in Fig.~\ref{fig:2}. The bulk effects are illustrated in Fig. \ref{fig:2}(a). When electrons and holes are generated by photoexcitation in the bulk, the photo-excited electrons possess a higher energy and have a higher probability of escaping as SEs after interacting with the PEs. This mechanism indicates that the photo-illuminated area should show a higher SEY and, thus, a ``bright'' contrast in the SUEM images. This mechanism, although not quantitatively examined so far, has been used to interpret the early SUEM results\cite{najafi2015four,liao2017photo}. In addition, the photogenerated bulk carriers  can also collide with SEs during their transport to the surface and prevent them from escaping the material surface, leading to a reduction of the SEY and a ``dark'' SUEM contrast. This mechanism was used to explain the dark SUEM contrasts observed on GaAs surfaces\cite{cho2014visualization}. Furthermore, the photoexcitation can also change the sample surface voltage and cause the surface electronic energy band to bend, known as the surface photovoltage effect (SPV) \cite{perovic1995field}. This effect is graphically illustrated in Fig.~\ref{fig:2}(b). On semiconductor surfaces, defects and dangling bonds can pin the surface Fermi level and lead to the bending of electronic bands near the surface that is associated with a surface electronic field. This surface field can facilitate or hinder the transmission of SEs across the sample surface, depending on the band bending direction. This effect is well known to cause different SEM image intensities from n-type and p-type surfaces\cite{perovic1995field}. Photoexcited photocarriers can compensate for the surface bending (the SPV effect), and thus modify the SE escape probability and the SEY. Fig.~\ref{fig:2}(b) shows the scenario in n-type silicon, where the surface bands bend upward near the surface under the dark condition, hindering the SE escape. In this case, SPV compensates for the surface band bending and lowers the surface barrier for SEs to escape, leading to a higher SEY and a bright SUEM contrast. Li et al. experimentally verified the impact of the SPV effect by observing a brighter (darker) SEM contrast in n-type (p-type) silicon under illumination compared to the dark state \cite{li2020probing}. The SPV effect at an internal interface was also recently studied using SUEM\cite{ellis2021scanning}.  Despite these previous efforts, there is no quantitative understanding of the relative contribution of these mechanisms to the SUEM contrast. This lack of understanding poses an obstacle to obtaining quantitative information from SUEM images.  Therefore, it is necessary to develop a computational approach to clarify the contributions from various contrast mechanisms. 

In this work, we implemented a Monte Carlo simulation assisted with time-dependent density functional theory (TDDFT) to quantitatively study the SUEM image contrast under photoexcitation with both bulk photocarrier effects and the surface SPV effect in the model system silicon and with PE energy from 50 eV to 1 keV. In particular, we used TDDFT to compute the effect of photoexcitation on the electron energy loss function (ELF) in silicon, and then used the ELF as input to a Monte Carlo simulation to evaluate the impact of photoexcitation on SEY. Our study laid the foundation for the future development of SUEM as a quantitative imaging tool.


\section{Methods}
\subsection{Monte Carlo Simulation}
Monte Carlo simulation is a well-established statistical technique for simulating the generation and interaction of SEs, backscattered electrons, and X-rays in electron microscopy and monitoring the trajectories and the properties of these particles, including the energy loss, mean free path, and escape probability\cite{joy1995monte,hovington1997casino,kuhr1999monte,ly1995monte,ding2005application}. Generally, Monte Carlo method is a step-wise simulation of electron trajectory by generating random numbers and predicting the scattering events based on the probability distribution from quantum mechanics. The simulated processes include tracing the trajectories of PEs, determining the position and results of scattering events, deciding the generation of SEs in the event of inelastic scattering and calculating the transmission probability of SEs at the sample surface. In the simulation, each simulated electron, either a PE or an SE, had a unique trajectory and experienced a separate sequence of scattering processes. By simulating a large number of trajectories, Monte Carlo simulation results can accurately match the experimental results and enable the analysis of the influence of multiple factors on the SEY. Our Monte Carlo simulation followed common procedures as detailed in \cite{ding1996a} and \cite{azzolini2017monte} and briefly summarized in the following paragraphs in this section.  

The main principle of the Monte Carlo method is to simulate electron scattering events by generating random numbers following particular distributions. All random numbers mentioned in this section followed a uniform distribution between 0 and 1. The path length traveled by one electron between two consecutive scatterings followed a Poisson distribution and was calculated by $ \Delta s=-\lambda_{\mathrm{tot}} \ln{r_1} $, where $\Delta s$ denoted the path length, $\lambda_{\mathrm{tot}}$ was the total mean free path, and $ r_1 $ represented the assigned random number \cite{azzolini2017monte}. The total mean free path was calculated through 
\begin{equation}
    \lambda_{\mathrm{tot}}=\frac{1}{\frac{1}{\lambda_{\mathrm{el}}}+\frac{1}{\lambda_{\mathrm{in}}}},
\end{equation}
where $\lambda_{\mathrm{el}}$ was the mean free path for elastic scatterings and $\lambda_{\mathrm{in}}$ was the mean free path for inelastic scatterings. 

The elastic mean free path was related to the elastic scattering cross section through
\begin{equation}
    \lambda_{\mathrm{el}}=\frac{1}{\sigma_{\mathrm{el}}n_{\mathrm{tot}}},
\end{equation}
where $\sigma_{\mathrm{el}}$ was the elastic scattering cross section and $n_{\mathrm{tot}}$ was the total number density of the atomic nuclei in the sample. The electronic elastic differential cross section in silicon as a function of the polar angle $\frac{d\sigma}{d\theta}$ was obtained through the NIST Electron Elastic-Scattering Cross-Section Database\cite{jablonski2004comparison}, which was based on the Mott theory of relativistic elastic scattering in a central field\cite{czyzewski1990calculations}. Since the elastic scatterings occur between electrons and the atomic nuclei, we assumed that the elastic scattering cross sections are not affected by photoexcitation, given the quantum states of the atomic nuclei are insensitive to low-energy visible photons.

Electrons can also undergo inelastic scatterings with other electrons inside the sample. During an inelastic scattering event, both the momentum and the energy of the participating electrons are altered. Since the inelastic scattering happens between electrons, the inelastic scattering mean free path $\lambda_{\mathrm{in}}$ should depend on both the local number density and the energy distribution of the electrons in the sample, which can be significantly changed by photoexcitation. Therefore, in order to evaluate how photoexcitation affects the SEY, it is necessary to quantitatively examine the impact of photoexcitation on the inelastic scattering mean free path. The inelastic scattering reflects the collective response of the ``electron sea'' inside the sample to an incident electron, and thus, can be described by the dielectric function $\varepsilon(\mathbf{q},\omega)$, where $\mathbf{q}$ is the momentum transfer during the scattering, and $\hbar \omega$ equals the energy loss $\Delta E$ of the incident electron due to the scattering\cite{pines1964elementary}. Within the linear response theory, the differential inelastic scattering mean free path can be connected to the dielectric function through\cite{pines1966the}
\begin{equation}
    \frac{d^2 \lambda_{\mathrm{in}}^{-1}}{d(\hbar \omega) dq}=\frac{1}{\pi a_0 E} \text{Im}  \left[ \frac{-1}{\varepsilon(q,\omega)} \right] \frac{1}{q},
    \label{eqn:inelastic_mfp}
\end{equation}
where $a_0$ is the Bohr radius and $E$ is the kinetic energy of the incident electron. Therefore, the probability for an inelastic scattering event, the energy loss, and the angular distribution after the inelastic scattering event are entirely encoded in the function $\text{Im}  [ \frac{-1}{\varepsilon(q,\omega)} ]$, which is termed the energy loss function (ELF). So our central task in this work is to evaluate how ELF is affected by photoexcitation, which we will detail in \ref{sec:ELF}. After the differential inelastic mean free path is evaluated from Eqn.~\ref{eqn:inelastic_mfp}, the total inelastic scattering mean free path can be obtained by integrating Eqn.~\ref{eqn:inelastic_mfp} over all possible energy losses and scattering angles. In a semiconductor, the minimum energy loss $\Delta E_{\mathrm{min}}$ is the band gap $E_G$, while the maximum energy loss $\Delta E_{\mathrm{max}}$, by convention, is half the kinetic energy of the incident electron. 

After the travel path length is determined, another random number $r_2$ is used to determine whether the collision is elastic or inelastic by comparing it to the ratio of the cross sections
\begin{equation}
    p_{\mathrm{el}}(E)=\frac{\sigma_{\mathrm{el}}(E)}{\sigma_{\mathrm{el}}(E)+\sigma_{\mathrm{in}}(E)}=\frac{1}{\lambda_{\mathrm{el}}(E)}\big(\frac{1}{\lambda_{\mathrm{el}}(E)}+\frac{1}{\lambda_{\mathrm{el}}(E)}\big)^{-1}.
\end{equation}
The collision is elastic when the random number is smaller than $p_{\mathrm{el}}$, and inelastic otherwise. Once the type of scattering is decided, other random numbers are assigned to determine the change in the polar angle $\theta$, the azimuthal angle $\phi$, and the energy loss if the scattering is inelastic. The change of the polar angle after both elastic and inelastic scatterings is determined by the corresponding differential cross sections. For elastic scatterings, the energy of the electron is conserved while its traveling direction is changed. The change in the polar angle $\theta$ after the scattering is determined by comparing the integrated differential cross section with a random number $r_3$:  
\begin{equation}
    \int_0^\theta \frac{d \sigma_{\mathrm{el}}}{d \Omega} \sin{\theta ^{\prime}} \, d\theta^{\prime}= r_3 \int_0^\pi \frac{d \sigma_{\mathrm{el}}}{d \Omega} \sin{\theta ^{\prime}} \, d\theta^{\prime},
    \label{eqn:elastic_angle}
\end{equation}
where $d\Omega=\sin{\theta} \, d\theta$ is the differential solid angle. As for inelastic scatterings, the energy loss $\Delta E$ after one inelastic scattering can be determined using another random number $r_4$:
 \begin{equation}
     \int_{\Delta E_{\mathrm{min}}}^{\Delta E} \frac{d \lambda_{\mathrm{in}}^{-1}}{d (\hbar \omega)} \, d(\hbar \omega) = r_4 \int_{\Delta E_{\mathrm{min}}}^{\Delta E_{\mathrm{max}}} \frac{d \lambda_{in}^{-1}}{d (\hbar \omega)} \, d(\hbar \omega).
 \end{equation}
 The polar angle change after an inelastic scattering can be determined in a similar way as in Eqn.~\ref{eqn:elastic_angle}, where the differential inelastic scattering cross section should be used. The change in the azimuthal angle $\phi$ after both elastic and inelastic scattering events is uniformly distributed between $0$ and $2 \pi$\cite{ding1996a}. Another consequence of each inelastic scattering is the generation of a secondary electron, whose kinetic energy $E_s$ equals the difference between the energy loss of the primary electron and the band gap energy:
 \begin{equation}
     E_s = \Delta E - E_G.
 \end{equation}
 The initial polar angle $\theta_s$ and the azimuthal angle $\phi_s$ of the secondary electron are related to the angle changes of the primary electron through momentum conservation\cite{ding1996a}:
 \begin{equation}
     \sin{\theta_s}=\cos{\theta}, ~\phi_s=\pi+\phi.
 \end{equation}
 Once the secondary electron is generated, its subsequent trajectory and scattering events will be tracked in the same way as described above, until it loses all of its kinetic energy or reaches the sample surface.

The status of each primary or secondary electron is checked after each scattering. Once one electron reaches the sample surface, its kinetic energy is compared with the surface energy barrier, the electron affinity $\chi$. If its kinetic energy is higher than the electron affinity $\chi$, the electron has a certain probability to escape the surface and contribute to the SEY. From elementary quantum mechanics, the transmission probability can be determined by the kinetic energy of the electron $E$, the incident angle $\alpha$, and the electron affinity $\chi$\cite{rao1974x}: \begin{equation} 
T=\frac{4\left(1-\frac{\chi}{E \cos \alpha}\right)^{1 / 2}}{\left[1+\left(1-\frac{\chi}{E \cos \alpha}\right)^{1 / 2}\right]^{2}}.
\label{eqn:transmission}
\end{equation} 
The electron is transmitted if a generated random number $r_8$ is smaller than the transmission rate, where the refraction angle satisfies $\sin \alpha^{\prime}=\left(\frac{E}{E-\chi}\right)^{1 / 2} \sin \alpha$, where $\alpha^{\prime}$ is the final escape angle into the vacuum \cite{rao1974x}. Otherwise, the electron is either terminated when its kinetic energy could not overcome the electron affinity or reflected back into the material and continues moving. Since the SPV effect will change the effective surface energy barrier for the escape of secondary electrons, we will discuss how the change of $\chi$ quantitatively affects the SEY in a later section. Finally, the SEY is calculated by repeating the above procedures for a large number of primary electrons. For this work, $10^7$ primary electrons were simulated in each case.

\subsection{Evaluating the Energy Loss Function}
\label{sec:ELF}
Many methods have been developed to calculate ELF. For example, ELF with zero momentum transfer $\text{Im}\left[ \frac{1}{\varepsilon(q=0,\omega)}\right]$ can be obtained from optical and electron energy loss spectroscopy (EELS) measurements and parameterized into the sum of a series of Drude-Lindhard-type terms\cite{ding1996a}:
\begin{equation}
    \textmd{Im}\left[\frac{-1}{\varepsilon(q=0,\omega)}\right] =\sum_{i=1}^{N} a_{i} \frac{\omega_{p i}^{2} \gamma_{i} \omega}{\left(\omega^{2}-\omega_{p i}^{2}\right)^{2}+\left(\gamma_{i} \omega\right)^{2}},
\end{equation}
where $\omega_{pi}$ is the frequency of the $i$-th plasmon peak, $\gamma_i$ is the associated damping coefficient, and the expansion coefficients $a_i$ are used to fit the experimental data. The $q=0$ ELF can then be extrapolated to finite momentum transfer by utilizing the quadratic plasmon dispersion relation:
\begin{equation}
    \omega_{pi}(q)=\omega_{pi}(q=0)+\frac{\hbar^2 q^2}{2m},
    \label{eqn:plasmon}
\end{equation}
where $m$ is the electron mass\cite{ashley1990energy}. Although this Drude-Lindhard (DL) model has been extensively used in Monte Carlo simulations of high energy electrons in solids, for our purpose, we need to compute the ELF from first principles in order to incorporate the photoexcitation effect. Later, we will use the DL model to benchmark our results.

In this work, we adopted time-dependent density functional theory (TDDFT) to evaluate the ELF from first principles\cite{ullrich2012time}. TDDFT has emerged as a tool to evaluate the dielectric properties of crystals, even in excited states\cite{weissker2006signatures,sato2014dielectric}. The principles of computing the dielectric function using TDDFT can be found elsewhere\cite{ullrich2012time} and are only briefly summarized here. First, the independent-particle polarizability $\chi^0_{\mathbf{G},\mathbf{G'}}(\mathbf{q},\omega)$ as a matrix in the basis of the plane waves labeled by the reciprocal lattice vectors $\mathbf{G}$ and $\mathbf{G'}$ can be calculated with the eigenvalues and wavefunctions obtained from solving Kohn-Sham equations in a standard DFT calculation\cite{hung2016interpretation}. Then, the TDDFT full polarizability $\chi_{\mathbf{G},\mathbf{G'}}(\mathbf{q},\omega)$ can be calculated through the Dyson equation:
\begin{equation}
    \chi_{\mathbf{G},\mathbf{G'}}(\mathbf{q},\omega) = \chi^0_{\mathbf{G},\mathbf{G'}}(\mathbf{q},\omega)+\chi^0_{\mathbf{G},\mathbf{G'}}(\mathbf{q},\omega)[v(\mathbf{q})+f_{xc}(\mathbf{q},\omega)]\chi_{\mathbf{G},\mathbf{G'}}(\mathbf{q},\omega),
\end{equation}
where $v(\mathbf{q})$ is the Coulomb potential and $f_{xc}(\mathbf{q},\omega)$ is the energy-dependent exchange-correlation kernel. In this work, we utilized a bootstrap exchange-correlation kernel\cite{sharma2011bootstrap} that includes effects beyond the random phase approximation. Next, the microscopic dielectric function $\varepsilon_{\mathbf{G},\mathbf{G'}}(\mathbf{q},\omega)$ is connected to the polarizability within the linear response theory\cite{azzolini2017monte}:
\begin{equation}
    \varepsilon^{-1}_{\mathbf{G},\mathbf{G'}}(\mathbf{q},\omega)=\delta_{\mathbf{G},\mathbf{G'}}+v(\mathbf{q})\chi_{\mathbf{G},\mathbf{G'}}(\mathbf{q},\omega).
\end{equation}
In systems where the local-field effect can be neglected, the macroscopic dielectric function $\epsilon(\mathbf{q},\omega)$ can be simply computed by:
\begin{equation}
    \varepsilon(\mathbf{q},\omega) = [\varepsilon^{-1}_{\mathbf{G}=\mathbf{0},\mathbf{G'}=\mathbf{0}}(\mathbf{q},\omega)]^{-1},
\end{equation}
from which the ELF can be obtained. Due to the heavy computational cost to calculate the full $\mathbf{q}$- and $\omega$-dependent ELF, we instead calculated the ELF with $q=0$ and extrapolated it to finite $q$ values using the plasmon dispersion relation (Eqn.~\ref{eqn:plasmon}), following the optical-data model by Ashley\cite{ashley1990energy}:
\begin{equation}
   \textmd{Im}\left\{-\frac{1}{\varepsilon(\mathbf{q}, \omega)}\right\}=\int_{0}^{\infty} \mathrm{d} \omega^{\prime}\left(\frac{\omega^{\prime}}{\omega}\right) \times \textmd{Im}\left\{-\frac{1}{\varepsilon\left(0, \omega^{\prime}\right)}\right\} \delta\left(\omega-\omega^{\prime}-\frac{\hbar}{2 m} \mathrm{q}^{2}\right).
   \label{eqn:ashley}
\end{equation}

We used the ELK code\cite{elk} for the TDDFT calculations in this work. Specifically, we used a local spin-density approximation to the exchange-correlation functional for the ground-state DFT and the bootstrap kernel $f_{xc}$ for the TDDFT\cite{sharma2011bootstrap}. A $k$-point sampling mesh of $15 \times 15 \times 15$ was used with a plane-wave cut-off energy of 200 eV. These parameters were tested to ensure the convergence of the results.

\begin{figure}
    \centering
    \includegraphics[width=0.95\textwidth]{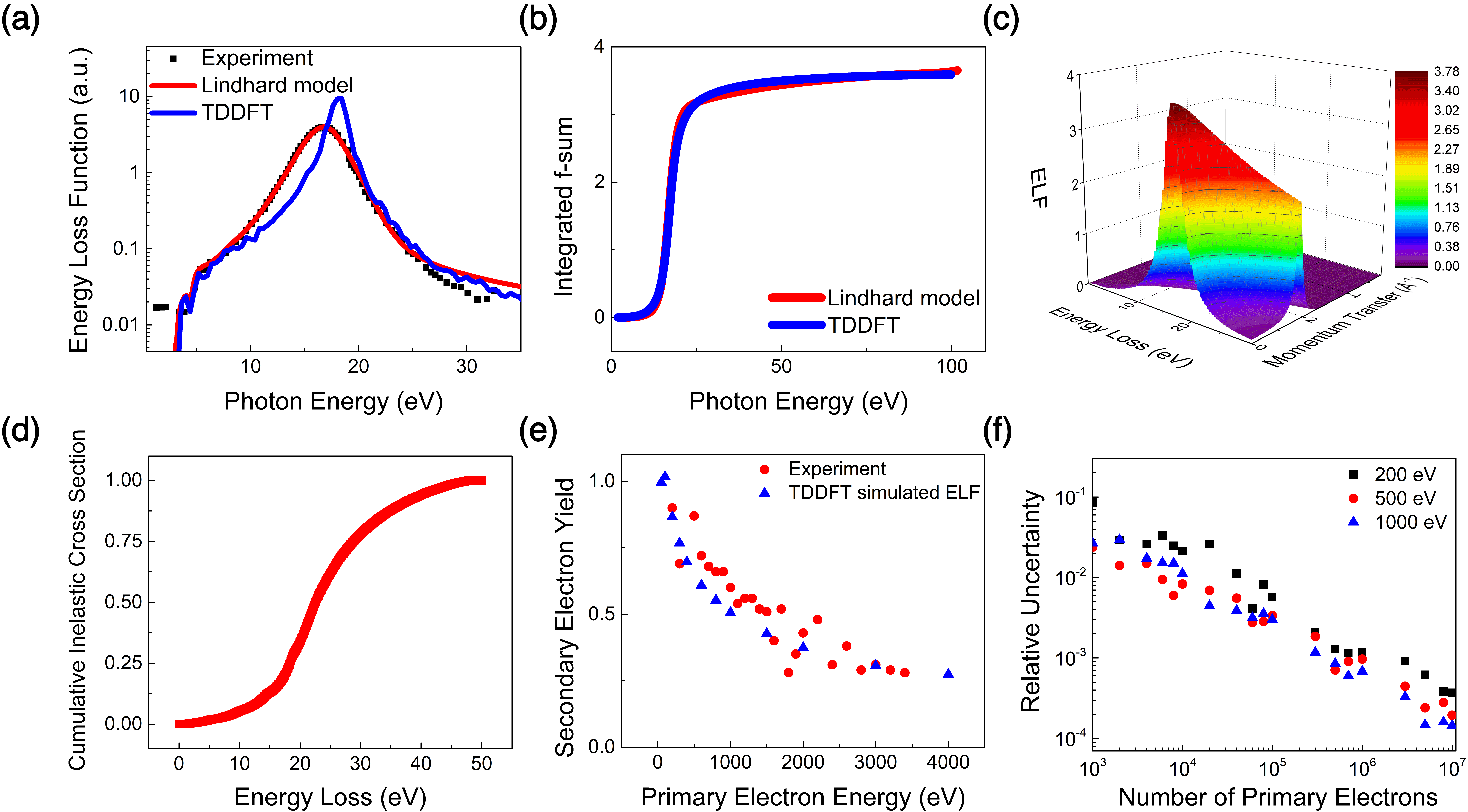}
    \caption{\textbf{Benchmarking the Monte Carlo Simulation.} (a) Comparing the ELF calculated using TDDFT to the experimental data and the Drude-Lindhard model. Experimental data is from Palik\cite{palik1998handbook}.(b) Checking the f-sum rule for the ELF. (c) The momentum- and energy-loss-dependent ELF after extrapolating the $q=0$ ELF calculated with TDDFT. (d) The cumulative inelastic scattering cross section calculated from the TDDFT ELF. (e) Comparing the SEY calculated using the TDDFT ELF to the experimental data. The experimental data is from Joy\cite{joy1995database}. (f) Scaling of the relative uncertainty of the Monte Carlo simulation as a function of the number of PEs simulated. 
}
    \label{fig:3}
\end{figure}

Our calculated zero-momentum ELF of silicon assuming an electron temperature of 300 K is shown in Fig.~\ref{fig:3}(a), in comparison to experimental data and the DL model. The experimental data is from the Handbook compiled by Palik\cite{palik1998handbook}, and the DL model parameters are from Sun et al.\cite{sun2016calculations}. From Fig.~\ref{fig:3}(a), the TDDFT calculation reproduces the low-energy-loss part of the ELF (below 10 eV) well while the position and the width of the bulk plasmon peak near 20 eV are slightly off from the experimental data measured by electron energy loss spectroscopy (EELS). This is most likely due to the fact that the electron temperature could be much higher than 300 K during the EELS measurement. The electron temperature effect will be discussed in detail in Sec.~\ref{sec:bulk}. Nevertheless, using the ELF calculated from TDDFT, we were able to obtain SEY from the Monte Carlo simulation in good agreement with experiments [Fig.~\ref{fig:3}(e)]. In addition, high-energy-loss channels associated with core electron ionizations (near 100 eV and 2000 eV in silicon\cite{sun2016calculations}) are not considered in this work since the core electrons are not influenced by photoexcitation in the visible range. In Fig.~\ref{fig:3}(b), we further test the ELF by checking the oscillator-strength sum rule (the ``f-sum'' rule)\cite{tanuma1993use} based on the Kramers-Kronig relation:
\begin{equation}
    Z_{\mathrm{eff}}=\frac{2}{\pi \Omega^2_{p}} \int_0^{\infty} \omega \textmd{Im} \left [ \frac{-1}{\varepsilon(\mathbf{q},\omega)} \right ] \, d\omega,
    \label{eqn:fsum}
\end{equation}
where $\Omega_p = \sqrt{4 \pi n_a}$ with $n_a$ being the atomic number density of the material, and $Z_{\mathrm{eff}}$ is the effective number of charge per atom. From Fig.~\ref{fig:3}(b), the f-sum rule integration up to 100 eV of our calculated TDDFT approaches $4$, the number of valence electrons per silicon atom, as expected. In Fig.~\ref{fig:3}(c), we show the momentum- and energy-loss-dependent ELF after extrapolating the $q=0$ ELF using Eqn.~\ref{eqn:ashley}. Figure ~\ref{fig:3}(d) shows the cumulative inelastic scattering cross section as a function of energy loss after scattering, demonstrating the major contribution of the bulk plasmon peak near 17 eV. Figure~\ref{fig:3}(e) compares the SEY generated using the Monte Carlo simulation with the ELF calculated from TDDFT to the experimental data compiled by Joy\cite{joy1995database} as a function of the incident PE energy. Reasonable agreement is achieved, considering the sizable scatter of the experimental data. Particularly, the peak SEY around 100-eV PE energy is reproduced. In Fig.~\ref{fig:3}(f), we further tested the statistical uncertainty of the SEY in the Monte Carlo simulation as a function of the number of independent PEs simulated. The statistical uncertainly is critical for our work since we anticipate that the change in SEY induced by photoexcitation to be quite small. As shown in Fig.~\ref{fig:3}(f), the relative statistical uncertainty in the SEY is reduced to below 0.1\% when $10^7$ PEs are simulated at different PE energies, defining the resolution of our Monte Carlo study. 

\section{Results and Discussion}
\subsection{Bulk Effect}
\label{sec:bulk}
As discussed above, photoexcitation can generate excess electrons and holes in the sample that can impact the SEY. As proposed in previous SUEM papers\cite{najafi2015four,liao2017scanning}, excess electrons (holes) increase (decrease) the local average electron energy, thus increasing (decreasing) the generation rate of SEs locally. In addition, excess electrons and holes inside the sample can scatter with the SEs inelastically during their transport inside the sample\cite{cho2014visualization}, thus reducing the number of SEs that can escape the sample surface. In principle, both effects are encoded in the ELF, which governs the SE generation and inelastic scattering processes. Furthermore, we focus on the bulk plasmon peak near 17 eV in silicon that is associated with its valence electrons that can be directly excited by the incident visible photons. In this section, we analyze how the ELF is affected by the presence of excess electrons and/or holes using TDDFT.

\begin{figure}[b]
    \centering
    \includegraphics[width=0.95\textwidth]{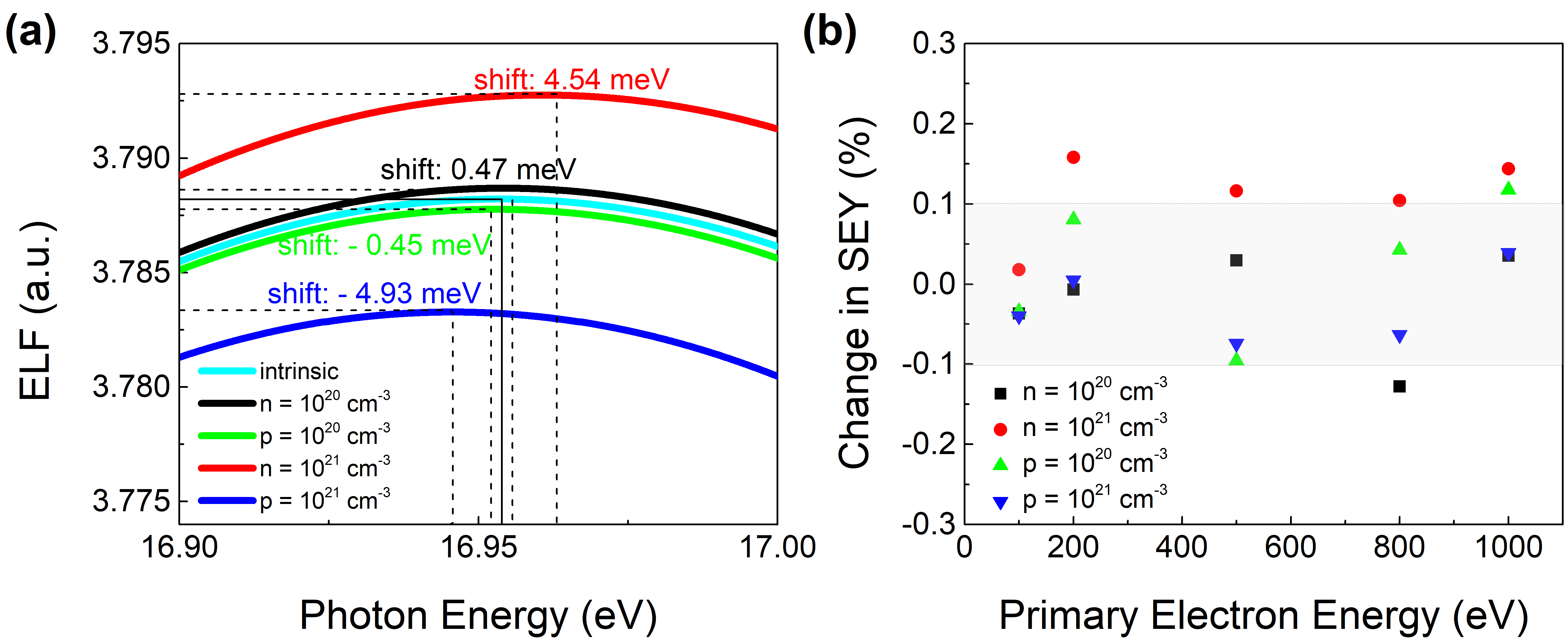}
    \caption{ \textbf{Impact of the Bulk Photocarriers on the Secondary Electron Yield.} (a) The ELF in the presence of excess electrons or holes calculated by TDDFT. The shift of the bulk plasmon peak is labeled in the figure. (b) The changes in the SEY from the Monte Carlo simulation taking into account the changes in the ELF. The gray zone labels the relative uncertainty in the Monte Carlo simulation with $10^7$ primary electrons simulated.}
    \label{fig:4}
\end{figure}

We simulated the effect of excess charge by adding extra electrons or holes in the TDDFT calculation. Considering practical optical excitation conditions\cite{sjodin1998ultrafast}, we simulated the excess carrier concentration up to $10^{21} ~\textmd{cm}^{-3}$ that is below the damage threshold of silicon. The calculated ELFs with different excess electron concentrations are shown in Fig.~\ref{fig:4}(a). A small shift of the plasmon peak position is observed. The plasmon frequency $\omega_p$ is proportional to $\sqrt{n_e}$, where $n_e$ is the density of valence electrons. In silicon, $n_e$ can be calculated to be $5 \times 10^{22} ~ \textmd{cm}^{-3}$. Therefore, a change in the valence electron density $\Delta n_e$ due to the presence of excess electrons and holes can lead to a change in the plasma frequency on the order of $\frac{\Delta n_e}{2 n_e}$. Thus, we expect a blue (red) shift of the plasmon peak by roughly 1\% when the excess electron (hole) concentration is $10^{21} ~ \textmd{cm}^{-3}$. This change in the ELF is small, mainly due to the fact that the concentration of excess charge that can be photoexcited is much lower in comparison to the valence electron concentration. In typical SUEM measurements with a pump photon energy of 2.4 eV, photoexcitation mainly generates excess electrons and holes near the band gap of the material, which will affect the inelastic scattering events on the energy scale of the band gap (1.1 eV in silicon). Since these scattering events occur at a much lower energy scale than the electron affinity (4.05 eV in silicon) that SEs need to overcome to escape, they are not expected to alter the SEY collected in SUEM experiments.

Although the change in the bulk plasmon peak due to photoexcitation is small (on the order of 0.1 eV and lower, depending on the photocarrier concentration), this small change can, in principle, be amplified due to the large number of inelastic scattering events experienced by each PE. Therefore, we still proceeded with Monte Carlo simulation and quantitatively evaluated the effect of the bulk photogenerated carriers on the SEY. In Fig.~\ref{fig:4}(b), we show the SEY from the Monte Carlo simulation with different excess electron or hole concentrations up to $10^{21}$ cm$^{-3}$ for the PE energy range below 1 keV. With the uncertainty level of $\sim$0.1\% achieved by simulating $10^7$ PEs, the changes in SEY caused by the photocarriers are comparable to the uncertainty level. Furthermore, considering the fact that electrons and holes coexist after photoexcitation and their excess concentrations cancel each other, we conclude here that the existence of photogenerated bulk carriers does not lead to sufficient changes in the SEY that can be detected within SUEM. This conclusion is consistent with a qualitative argument provided by Zhang et al.\cite{zhang2021photoabsorption}, and we provided the numerical proof here.

\begin{figure}[b]
    \centering
    \includegraphics[width=0.95\textwidth]{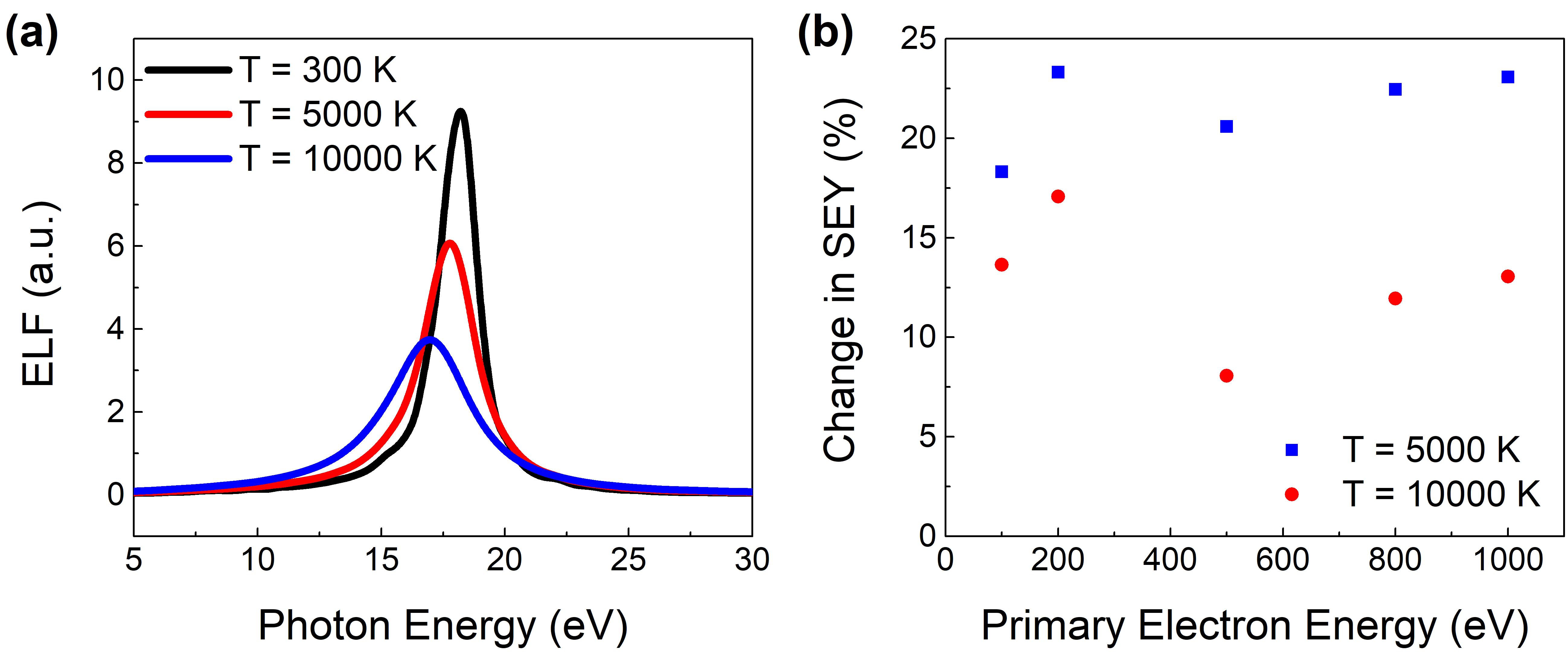}
    \caption{ \textbf{Impact of Hot Photocarrier Effect on the Secondary Electron Yield.} (a) The ELF calculated by TDDFT with different effective electron temperatures. (b) The change in SEY due to different effective electron temperatures (compared to that at 300 K) from the Monte Carlo simulation as a function of the PE energy.}
    \label{fig:5}
\end{figure}

The analysis in this section so far has assumed that the photogenerated bulk carriers are in thermal equilibrium with the lattice. In practice, however, immediately after the excitation by a short optical pulse, the photogenerated electrons and holes can possess a high effective temperature depending on the frequency of the optical pulse used:
\begin{equation}
    k_B T = h \nu-E_G,
\end{equation}
where $T$ is the effective charge carrier temperature, $k_B$ is the Boltzmann constant and $\nu$ is the frequency of the optical pulse. With an optical pulse in the visible range, the effective temperature of the photogenerated ``hot'' electrons and holes can be as high as tens of thousands of Kelvin\cite{Gitomer_Jones_Begay_Ehler_Kephart_Kristal_1986}. Since the initial high temperature will lead to significant broadening of the electron distribution in the bulk on the order of a few eV, we expect this hot photocarrier effect can significantly alter ELF and influence the SEY. In fact, the hot carrier transport effect has been observed in SUEM experiments in a wide range of materials\cite{liao2017photo,Najafi2017,el2019extraordinary,choudhry2022persistent}. Here, we simulated the hot carrier effect by adjusting the smearing width of the equilibrium electron distribution function in the TDDFT calculation of the ELF. As a first-order approximation, we set the smearing width of the equilibrium electron distribution function to be $k_B T$. In Fig.~\ref{fig:5}(a), we compare the ELF calculated with an effective electron temperature at 300 K, 5000 K, and 10000 K, respectively. As expected, a higher effective electron temperature broadens the bulk plasmon peak in the ELF while suppressing its peak value, as required by the f-sum rule, since the total number of valence charge remains the same at different electron temperatures (in fact, the ELF calculated with an electron temperature of 10000 K reproduces the position and the width of the bulk plasmon peak measured experimentally by EELS, as shown in Fig.~\ref{fig:3}(a), suggesting the electrons could be in a highly nonequilibrium state during the EELS measurement). This indicates that a higher effective electron temperature will lead to excitation of SEs with a broader energy distribution near the bulk plasmon peak. In the process, more high-energy SEs can be created that can give rise to an increased SEY. To quantitatively evaluate this effect, we conducted Monte Carlo simulation with the ELFs calculated at different electron temperatures. The result is shown in Fig.~\ref{fig:5}(b). Within the range of PE energies simulated in this work (100 to 1000 eV), the change in SEY due to a high effective electron temperature is quite significant. The largest increase in SEY is about 23\% at 5000 K for a PE energy of 200 eV. Interestingly, SEY does not increase with temperature monotonically, since the SEY is higher at 5000 K than that at 10000 K. This can be understood since the broadening of the ELF with increasing temperature is also accompanied by a shift of the bulk plasmon peak towards lower energy, creating an opposite impact on SEY. This result suggests that the hot carrier effect right after photoexcitation can generate a strong contrast in SUEM experiments, while the presence of excess electrons and holes in bulk of the sample after they thermally equilibrate with the environment has a negligible effect on the SUEM contrast.

\subsection{Surface effect}
\begin{figure}[b]
    \centering
    \includegraphics[width=0.95\textwidth]{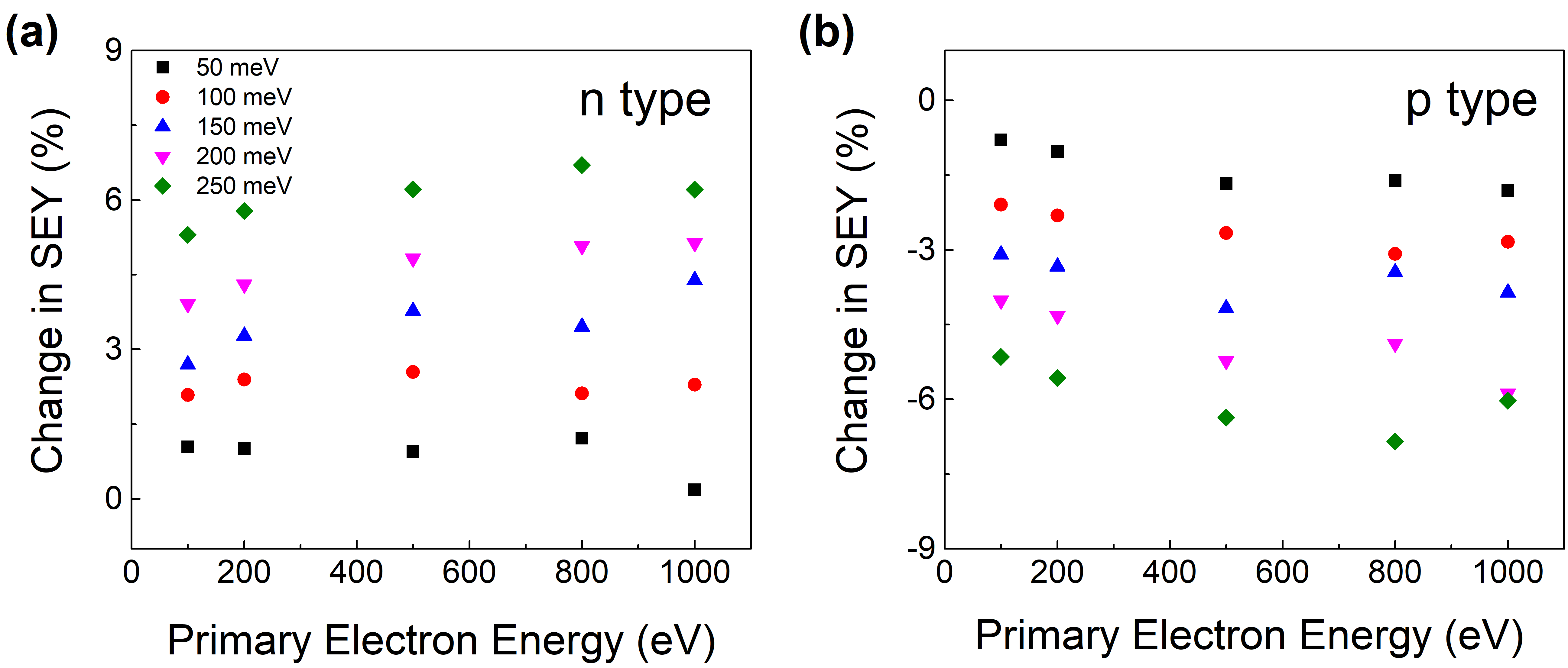}
    \caption{\textbf{Impact of the Surface Photovoltage Effect on the Secondary Electron Yield.} The change in the SEY as the effective surface potential $\chi$ is changed by different amounts, up to 250 meV, in (a) n-type silicon and (b) p-type silicon, as a function of the PE energy.}
    \label{fig:6}
\end{figure}

Besides the bulk carrier effects, photoexcitation can also induce the SPV effect, which modifies the surface transmission barrier for the SE escape that impacts the SEY. As explained in the Introduction section, the SPV was originated from the surface band bending in semiconductors due to the Fermi level pinning by surface defect states \cite{Zhang_Yates_2012}, which can be compensated by the photoexcited charge carriers \cite{li2020probing}. In effect, the SPV changes the effective electron affinity $\chi$ in Eqn.~\ref{eqn:transmission}, and thus, can be simulated by modifying the $\chi$ parameter in the Monte Carlo simulation. Zhang et al.\cite{zhang2021photoabsorption} conducted a similar simulation of spherical nanoparticles. Our simulation results of a flat silicon surface are shown in Fig.~\ref{fig:6}. In n-type silicon, the SPV effect leads to a reduction of the effective SE escape barrier, and thus, an increased SEY. The opposite effect occurs in p-type silicon. In both cases, we simulated an SPV voltage up to 250 meV, as has been measured experimentally in silicon\cite{li2020probing}. For a given PE energy, the change in the SEY scales monotonically with the the SPV voltage in both n-type and p-type silicon. The change in SEY increases with increasing PE energy initially and seems to saturate around 1 keV PE energy. The change in SEY due to SPV is on the order of a few percent in our simulation, in agreement with the previous report by Zhang et al.\cite{zhang2021photoabsorption}. This result indicates that once the photoexcited hot carriers cool down, the SPV effect plays a major role in determining the SUEM contrast, rather than the distribution of the bulk photocarriers. 

\section{Conclusions}
In summary, we quantitatively examined the impact of photoexcitation on the SEY by combining Monte Carlo simulation with an ELF calculated from TDDFT. The effect of both photoexcited carriers in the bulk and the altered surface potential is evaluated. We found that the hot carrier effect immediately after photoexcitation can have a strong impact on the SEY, while the mere presence of bulk photocarriers after they cool down leads to a negligible change in SEY. In addition, the SPV effect induced by the photocarriers near the sample surface adds another contrast mechanism to SUEM images. Due to the limit in computational resources, we only focused our study on low energy PEs below 1 keV and on silicon as a model system, so we caution against direct quantitative comparison to SUEM experiments with high PE energies (typically 30 keV) and in other materials. However, we believe the qualitative observations from this study still provide important physical insights into understanding the SUEM contrast, and we will follow up with simulations of higher PE energies and other materials in our future work.  

\section*{Conflict of Interest}
The authors have no conflicts to disclose.

\begin{acknowledgments}
We thank Professor Vojt\v{e}ch Vlc\v{e}k for helpful discussions. This work is based on research partially supported by the U.S. Department of Energy, Office of Basic Energy Sciences, under the award number DE-SC0019244 for the development of SUEM, and National Science Foundation under the award number DMR-1905389 for developing the Monte Carlo simulation. This work used Stampede2 at Texas Advanced Computing Center (TACC) through allocation MAT200011 from the Advanced Cyberinfrastructure Coordination Ecosystem: Services \& Support (ACCESS) program, which is supported by National Science Foundation grants 2138259, 2138286, 2138307, 2137603, and 2138296. Use was also made of computational facilities purchased with funds from the National Science Foundation (CNS-1725797) and administered by the Center for Scientific Computing (CSC) at UC Santa Barbara. The CSC is supported by the California NanoSystems Institute and the Materials Research Science and Engineering Center (MRSEC; NSF DMR-1720256) at UC Santa Barbara.
\end{acknowledgments}



\bibliography{references.bib}

\end{document}